# Study of Neural Network Algorithm for Straight-Line Drawings of Planar Graphs

Mohamed A. El-Sayed[a], S. Abdel-Khalek[b], and Hanan H. Amin[c]

[a] Mathematics department, Faculty of Science, Fayoum University, 63514 Fayoum, Egypt
[b,c] Mathematics department, Faculty of Science, Sohag University, 82524 Sohag, Egypt
[a] CS department, College of Computers and Information Technology, Taif Univesity, 21974 Taif, KSA
[b] Mathematics department, Faculty of Science, Taif Univesity, 21974 Taif, KSA
[a] drmasayed@yahoo.com , [b] abotalb2010@yahoo.com, [c] hananhamed85@yahoo.com

*Abstract*— Graph drawing addresses the problem of finding a layout of a graph that satisfies given aesthetic and understandability objectives. The most important objective in graph drawing is minimization of the number of crossings in the drawing, as the aesthetics and readability of graph drawings depend on the number of edge crossings. VLSI layouts with fewer crossings are more easily realizable and consequently cheaper. A straight-line drawing of a planar graph G of n vertices is a drawing of G such that each edge is drawn as a straight-line segment without edge crossings.
However, a problem with current graph layout methods which are capable of producing satisfactory results for a wide range of graphs is that they often put an extremely high demand on computational resources. This paper introduces a new layout method, which nicely draws internally convex of planar graph that consumes only little computational resources and does not need any heavy duty preprocessing. Here, we use two methods: The first is self organizing map known from unsupervised neural networks which is known as (SOM) and the second method is Inverse Self Organized Map (ISOM).

## I. INTRODUCTION

The drawing of graphs is widely recognized as a very important task in diverse fields of research and development. Examples include VLSI design, plant layout, software engineering and bioinformatics [13]. Large and complex graphs are natural ways of describing real world systems that involve interactions between objects: persons and/or organizations in social networks, articles incitation networks, web sites on the World Wide Web, proteins in regulatory networks, etc [23,10].

Graphs that can be drawn without edge crossings (i.e. planar graphs) have a natural advantage for visualization [12]. When we want to draw a graph to make the information contained in its structure easily accessible, it is highly desirable to have a drawing with as few edge crossings as possible.

A straight-line embedding of a plane graph G is a plane embedding of G in which edges are represented by straight-line segments joining their vertices, these straight line segments intersect only at a common vertex.

A straight-line drawing is called a convex drawing if every facial cycle is drawn as a convex polygon. Note that not all planar graphs admit a convex drawing. A straight-line drawing is called an inner-convex drawing if every inner facial cycle is drawn as a convex polygon.

A strictly convex drawing of a planar graph is a drawing with straight edges in which all faces, including the outer face, are strictly convex polygons, i. e., polygons whose interior angles are less than 180. [1]

However, a problem with current graph layout methods which are capable of producing satisfactory results for a wide range of graphs is that they often put an extremely high demand on computational resources [20].

One of the most popular drawing conventions is the straight-line drawing, where all the edges of a graph are drawn as straight-line segments. Every planar graph is known to have a planar straight-line drawing [8]. A straight-line drawing is called a convex drawing if every facial cycle is drawn as a convex polygon. Note that not all planar graphs admit a convex drawing. Tutte [25] gave a necessary and suifcient condition for a triconnected plane graph to admit a convex drawing. Thomassen [24] also gave a necessary and su.cient condition for a biconnected plane graph to admit a convex drawing. Based on Thomassen's result, Chiba et al. [6] presented a linear time algorithm for finding a convex drawing (if any) for a biconnected plane graph with a specified convex boundary. Tutte [25] also showed that every triconnected plane graph with a given boundary drawn as a convex polygon admits a convex drawing using the polygonal boundary. That is, when the vertices on the boundary are placed on a convex polygon, inner vertices can be placed on suitable positions so that each inner facial cycle forms a convex polygon.

In paper [15], it was proved that every triconnected plane graph admits an inner-convex drawing if its boundary is fixed with a star-shaped polygon P, i.e., a polygon P whose kernel (the set of all points from which all points in P are visible) is not empty. Note that this is an extension of the classical result by Tutte [25] since any convex polygon is a star-shaped polygon. We also presented a linear time algorithm for computing an inner-convex drawing of a triconnected plane graph with a star-shaped boundary [15].

This paper introduces layout methods, which nicely draws internally convex of planar graph that consumes only little computational resources and does not need any heavy duty preprocessing. Unlike other declarative layout algorithms not even the costly repeated evaluation of an objective function is required. Here, we use two methods: The first is self organizing map known from unsupervised neural networks which is known as (SOM) and the second method is Inverse Self Organized map (ISOM).

## II. PRELIMINARIES

Throughout the paper, a graph stands for a simple undirected graph unless stated otherwise. Let $G = (V,E)$ be a graph. The set of edges incident to a vertex $v \in V$ is denoted by $E(v)$. A vertex (respectively, a pair of vertices) in a connected graph is called a *cut vertex* (respectively, a *cut pair*) if its removal from $G$ results in a disconnected graph. A connected graph is called *biconnected* (respectively, *triconnected*) if it is simple and has no cut vertex (respectively, no cut pair).

We say that a cut pair *{u, v} separates* two vertices *s* and *t* if *s* and *t* belong to different components in *G-{u, v}*.

A graph $G = (V,E)$ is called *planar* if its vertices and edges are drawn as points and curves in the plane so that no two curves intersect except at their endpoints, where no two vertices are drawn at the same point. In such a drawing, the plane is divided into several connected regions, each of which is called a *face*. A face is characterized by the cycle of $G$ that surrounds the region. Such a cycle is called a *facial cycle*. A set $F$ of facial cycles in a drawing is called an *embedding* of a planar graph $G$.

A *plane* graph $G = (V, E, F)$ is a planar graph $G = (V,E)$ with a fixed embedding $F$ of $G$, where we always denote the outer facial cycle in $F$ by $f_o \in F$. A vertex (respectively, an edge) in $f_o$ is called an *outer vertex* (respectively, an *outer edge*), while a vertex (respectively, an edge) not in $f_o$ is called an *inner vertex* (respectively, an *inner edge*).

The set of vertices, set of edges and set of facial cycles of a plane graph $G$ may be denoted by $V(G)$, $E(G)$ and $F(G)$, respectively.

A biconnected plane graph $G$ is called *internally triconnected* if, for any cut pair *{u, v}*, *u* and *v* are outer vertices and each component in $G - \{u, v\}$ contains an outer vertex. Note that every inner vertex in an internally triconnected plane graph must be of degree at least 3.

A graph $G$ is *connected* if for every pair *{u, v}* of distinct vertices there is a path between *u* and *v*. The *connectivity* $\kappa(G)$ of a graph $G$ is the minimum number of vertices whose removal results in a disconnected graph or a single-vertex graph $K_1$. We say that $G$ is *k-connected* if $\kappa(G) \geq k$. In other words, a graph $G$ is *3-connected* if for any two vertices in $G$ are joined by three vertex-disjoint paths.

Define a plane graph $G$ to be *internally 3-connected* if (a) $G$ is 2-connected, and (b) if removing two vertices *u,v* disconnects $G$ then *u, v* belong to the outer face and each connected component of *G-{u, v}* has a vertex of the outer face. In other words, $G$ is internally 3-connected if and only if it can be extended to a 3-connected graph by adding a vertex and connecting it to all vertices on the outer face. Let $G$ be an n-vertex 3-connected plane graph with an edge $e(v_1,v_2)$ on the outer face.

## III. PREVIOUS WORKS IN NEURAL NETWORKS

Artificial neural networks have quite long history. The story has started with the work of W. McCulloch and W. Pitts in 1943 [21]. Their paper presented the first artificial computing model after the discovery of the biological neuron cell in the early years of the twentieth century. The McCulloch-Pitts paper was followed by the publication from F. Rosenblatt in 1953, in which he focused on the mathematics of the new discipline [22]. His perceptron model was extended by two famous scientists in [2].

The year 1961 brought the description of competitive learning and learning matrix by K. Steinbruch [5]. He published the "winner-takes-all" rule, which is widely used also in modern systems. C. von der Malsburg wrote a paper about the biological self-organization with strong mathematical connections [19]. The most known scientist is T. Kohonen associative and correlation matrix memories, and – of course – self-organizing (feature) maps (SOFM or SOM) [16,17,18]. This neuron model has great impact on the whole spectrum of informatics: from the linguistic applications to the data mining.

The Kohonen's neuron model is commonly used in different classification applications, such as the unsupervised clustering of remotely sensed images.

In NN it is important to distinguish between supervised and unsupervised learning. Supervised learning requires an external "teacher" and enables a network to perform according to some predefined objective function. Unsupervised learning, on the other hand, does not require a teacher or a known objective function: The net has to discover the optimization criteria itself. For the unsupervised layout task at hand this means that we will not use an objective function prescribing the layout aesthetics. Instead we will let the net discover these criteria itself. The best-known NN models of unsupervised learning are Hebbian learning [14] and the models of competitive learning: The adaptive resonance theory [10], and the self-organizing map or Kohonen network which will be illustrated in the following section

The basic idea of competitive learning is that a number of units compete for being the "winner" for a given input signal. This winner is the unit to be adapted such that it responds even better to this signal. In a NN typically the unit with the highest response is selected as the winner[20].

M. Hagenbuchner, A.Sperduti and A.C.Tsoi described a novel concept on the processing of graph structured information using the self-organizing map framework which allows the processing of much more general types of graphs, e.g. cyclic graphs [11] . The novel concept proposed in those paper, namely, by using the clusters formed in the state space of the self-organizing map to represent the ''strengths'' of the activation of the neighboring vertices. Such an approach resulted in reduced computational demand, and in allowing the processing of non-positional graphs.

Georg PÄolzlbauer, Andreas Rauber, Michael Dittenbach presented two novel techniques that take the density of the data into account. Our methods define graphs resulting from nearest neighbor- and radius-based distance calculations in data space and show projections of these graph structures on the map. It can then be observed how relations between the data are preserved by the projection, yielding interesting insights into the topology of the mapping, and helping to identify outliers as well as dense regions [9].

Bernd Meyer introduced a new layout method that consumes only little computational resources and does not need any heavy duty preprocessing. Unlike other declarative layout algorithms not even the costly repeated evaluation of an objective function is required. The method presented is based on a competitive learning algorithm which is an extension of self-organization strategies known from unsupervised neural networks[20].

## IV. SELF-ORGANIZING FEATURE MAPS ALGORITHM

Self-Organizing Feature Maps (SOFM or SOM) also known as Kohonen maps or topographic maps were first introduced by von der Malsburg [19] and in its present form by Kohonen [16].

According to Kohonen the idea of feature map formation can be stated as follows: The spatial location of an output neuron in the topographic map corresponds to a particular domain, or feature of the input data.

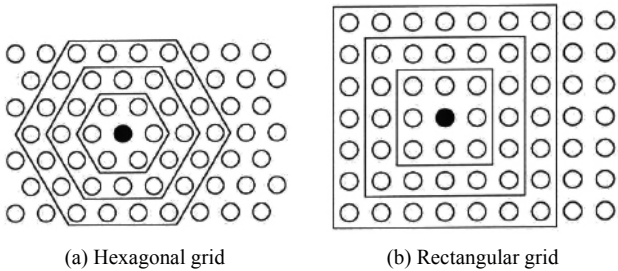

(a) Hexagonal grid        (b) Rectangular grid

Figure 1. rectangular and hexagonal 2- dimensional grid

The general structure of SOM or the Kohonen neural network which consists of an input layer and an output layer. The output layer is formed of neurons located on a regular 1- or 2- dimensional grid. In the case of the 2- dimensional grid, the neurons of the map can exist in a rectangular or a hexagonal topology, implying 8-neighborhood or 6 neighborhoods, respectively. as shown in Figure (1).

The network structure is a single layer of output units without lateral connections and a layer of n input units. Each of the output units is connected to each input unit.

Kohonen's learning procedure can be formulated as:
- Randomly present a stimulus vector $x$ to the network
- Determine the "winning" output node $u_i$, where $w_i$ is the weight vector connecting the inputs to output node $i$.

$$\|w_i - x\| \leq \|w_j - x\| \forall k$$

Note: the above equation is equivalent to $w_i.x >= w_j.x$ only if the weights are normalized.
- Given the winning node i, and adapt the weights of $w_k$ and all nodes in a neighborhood of a certain radius $r$, according to the function

$$w_i(new) = w_i(old) + \alpha.\Omega(u_i,u_j)(x - w_i)$$

- After every $j$-th stimulus decrease the radius $r$ and $\alpha$.

Where $\alpha$ is adaption factor and $\Omega(u_i,u_j)$ is a neighborhood function whose value decreases with increasing topological distance between $u_i$ and $u_j$.

The above rule drags the weight vector $w_i$ and the weights of nearby units towards the input $x$.

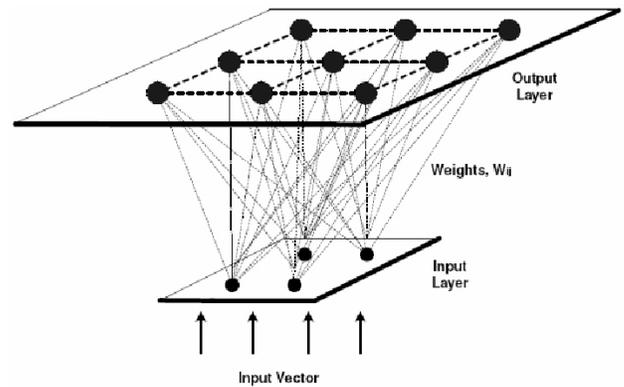

Figure 2. General structure of Kohonen neural network

This process is iterated until the learning rate α falls below a certain threshold. In fact, it is not necessary to compute the units' responses at all in order to find the winner. As Kohonen shows, we can as well select the winner unit $u_j$ to be the one with the smallest distance $\|\vec{v} - \vec{w_j}\|$ to the stimulus vector. In terms of Figure 3 this means that the weight vector of the winning unit is turned towards the current input vector.

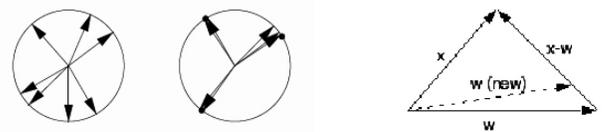

Figure 3. Adjusting the Weights.

Kohonen demonstrates impressively that for a suitable choice of the learning parameters the output network organizes itself as a topographic map of the input. Various forms are possible for these parameter functions, but negative exponential functions produce the best results, the intuition being that a coarse organization of the network is quickly achieved in early phases, whereas a localized fine organization is performed more slowly in later phases. Therefore common choices are: Gaussian neighborhood function $\Omega(u_i,u_j) = e^{-d(u_i,u_j)^2/2\sigma(t)^2}$ where $d(u_i,u_j)$ is the topological distance of $u_i$ and $u_j$ and $\sigma^2$ is the neighborhood width parameter that can gradually be decreased over time.

To get amore intuitive view of what is happening, we can now switch our attention to the weight space of the network. If we restrict the input to two dimensions, each weight vector can be interpreted as a position in two-dimensional space. Depicting the 4-neighborhood relation as straight lines between neighbors, Figure 4 illustrates the adaption process. Starting with the random distribution of weights on the left-hand side and using nine distinct random input stimuli at the positions marked by the black dots, the net will eventually settle into the organized topographic map on the right-hand side, where the units have moved to the positions of the input stimuli.

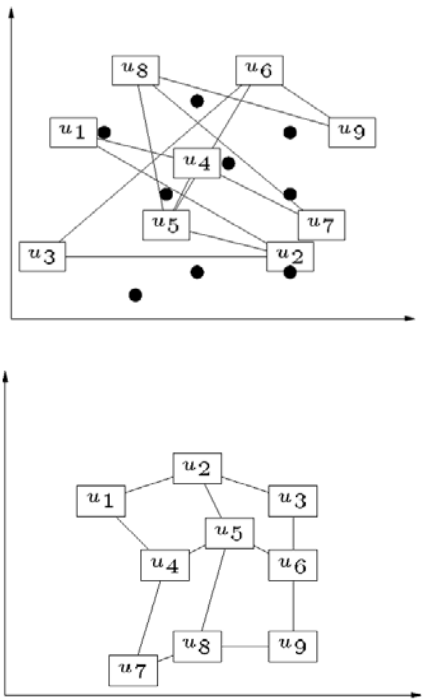

Figure 4. A Simple of random distribution of *G* and its the organized topographic map.

The SOM algorithm is controlled by two parameters: a factor $\alpha$ in the range 0…1, and a radius *r*, both of which decrease with time. We have found that the algorithm works well if the main loop is repeated 1,000,000 times. The algorithm begins with each node assigned to a random position. At each step of the algorithm, we choose a random point within the region that we want the network to cover ( rectangle or hexagonal), and find the closest node (in terms of Euclidean distance) to that point. We then move that node towards the random point by the fraction α of the distance. We also move nearby nodes (those with conceptual distance within the radius *r*) by a lesser amount [11,20].

The above SOM algorithm can be written as the following:
  input: An internally convex of planar graph $G=(V,E)$
  output: Embedding of a planar graph *G*
  radius $r := r_{max}$; /* initial radius */
  initial learning rate $\alpha_{max}$ ;
  final learning rate $\alpha_{min}$
  repeat many times
    choose random (*x,y*);
    *i* = index of closest node;
    move node *i* towards (*x,y*) by $\alpha$ ;
    move nodes with *d*<*r* towards (*x,y*) by $\alpha.e^{-d^2/2\sigma(t)^2}$ .
    decrease $\alpha$ and *r*;
  end repeat

## V. INVERTING THE SOM ALGORITHM (ISOM)

We can now detail the ISOM algorithm. Apart from the different treatment of network topology and input stimuli closely resembles Kohonen's method [20].

In ISOM there are Input layer and weights layer only the actual network output layer is discarded completely in this method we look at the weight space instead of at the output response and to interpret the weight space as a set of positions in space.

The main differences to the original SOM are not so much to be sought in the actual process of computation as interpretation of input and output. First, the problem input given to our method is the network topology and not the set of stimuli. The stimuli themselves are no longer part of the problem description as SOM but a fixed part of the algorithm, we are not really using the input stimuli at all, but we are using a fixed uniform distribution. For this reason, the layout model presented here will be called the inverted self-organizing map (ISOM). Secondly, we are interpreting the weight space as the output parameter.

In this method, there is no activation function σ in difference of SOM. In ISOM we use a parameter called "cooling" (c) and we use different decay or neighboring function: In the SOM method we use the neighborhood function $\Omega(u_i,u_j) = e^{-d(u_i,u_j)^2/2\sigma(t)^2}$ where $d(u_i,u_j)$ is the topological distance of $u_i$ and $u_j$ and $\sigma^2$ is the width parameter that can gradually be decreased over time .

In ISOM we use the neighborhood function $\Omega(u_i,u_j) = -2^{-d(w_i,w_j)}$ , where $d(w_i,w_j)$ is the distance between *w* and all successors $w_i$ of *w*.

The above ISOM algorithm can be written as the following:
  input: An internally convex of planar graph $G=(V,E)$
  output: Embedding of a planar graph *G*
  epoch $t := 1$;
  radius $r := r_{max}$; /* initial radius */
  initial learning rate $\alpha_{max}$ ;

```
cooling factor c;
forall v ∈ V do v.pos := random_vector();
while (t ≤ t_max) do
    Adaption α := max(min_adaption,…
                    e^{-c(t/t_max)} .max_adaption)
    i := random_vector();
    /* uniformly distributed in input area */
    w := v ∈ V such that ||v.pos − i⃗|| is minimal
    for w and all successors w_i of w with d(w,w_i) ≤ r :
    w_i.pos = w_i.pos − 2^{-d(w_i,w_j)}.α(w_i.pos − i);
    t:=t+1;
    if r > min_radius  do   r:=r-1;
end while.
```

The node positions $w_i.pos$ which take the role of the weights in the SOM are given by vectors so that the corresponding operations are vector operations. Also note the presence of a few extra parameters such as the minimal and maximal adaption, the minimal and initial radius, the cooling factor, and the maximum number of iterations. Good values for these parameters have to be found experimentally [20].

## VI. EXPERIMENTS AND RESULTS

The sequential algorithm of the SOM model and ISOM were designed in Matlab language for tests. The program runs on the platform of a GIGABYTE desktop with Intel Pentium (R) Dual-core CPU 3GHZ, and 2 GB RAM.

The algorithm was tested on randomly generated graphs $G=(V,E)$. Initially, all vertices are randomly distributed in this area grid unit, and the weights generate at random distribution points .The initial graph has been drawing by many crossing edges see figure (5.a) where the grid size is (4*4) nodes.

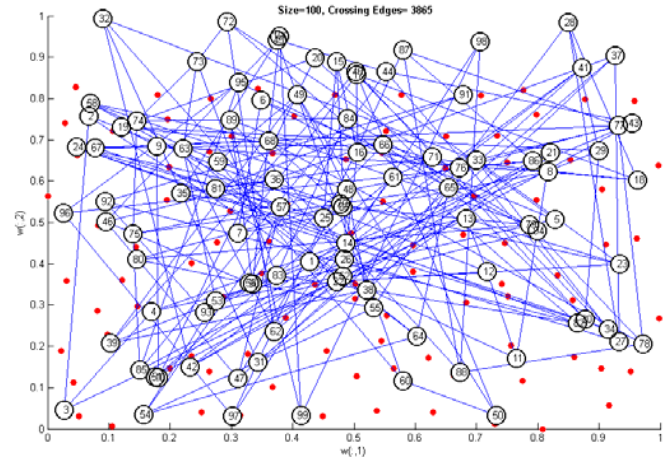

(a) random weights of *G*, size=100 node , edge crossing = 3865

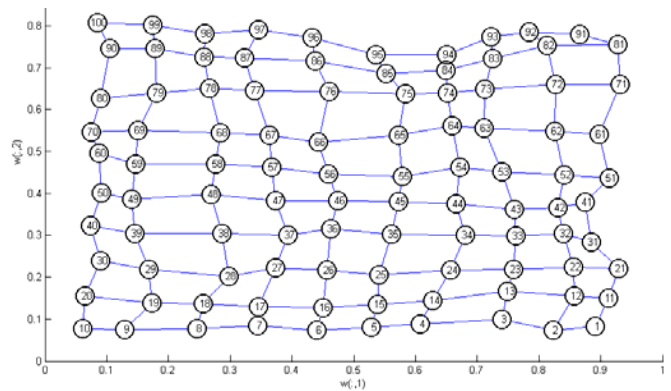

(b) SOM

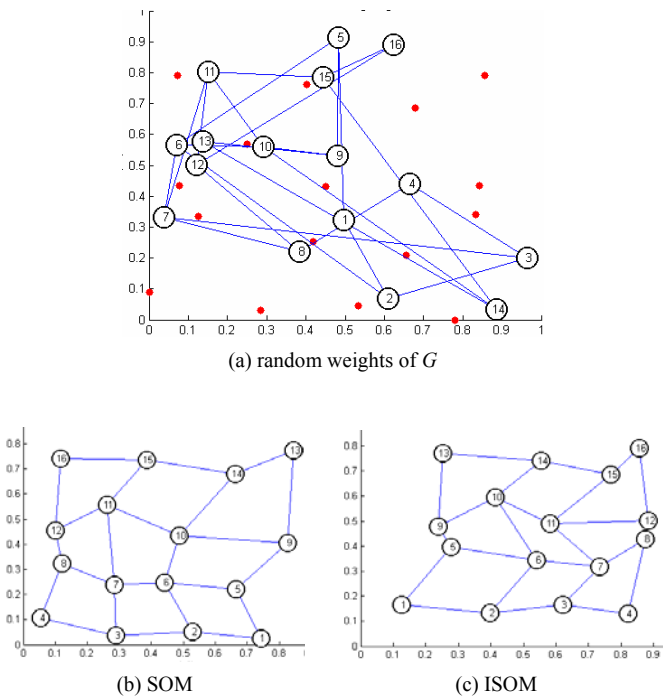

(a) random weights of *G*

(b) SOM          (c) ISOM

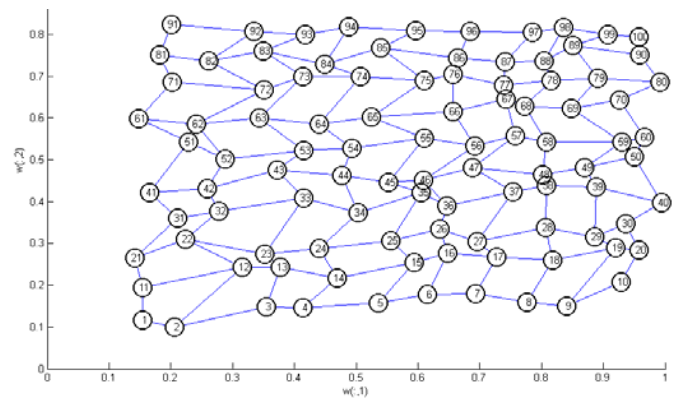

(c) ISOM

Figure 5.  random weights of graph with 16 nodes, output graph drawing using SOM and ISOM, respectively.

Figure 6.  random weights of graph with 100 nodes, output graph drawing using SOM and ISOM, respectively.

In the SOM method: The algorithm is controlled by two parameters: a factor α in the range 0…1, (we used initial learning rate at α=0.5 and the final at α=0.1) and a radius *r*, (the initial radius at 3) both of which decrease with time.

In the ISOM method: The choice of parameters can be important. However the algorithm seems fairly robust against small parameter changes and the network usually quickly settles into one of a few stable configurations. As a rule of thumb for medium sized graphs, 1000 epochs with a cooling factor *c*=1.0 yield good results. The initial radius obviously depends on the size and connectivity of the graph and initial radius *r*=3 with an initial adaption of 0.8 was used for the examples in our paper. It is important that the intervals for radius and adaption both of which decrease with time. The final phase with *r*=0 should only use very small adaption factors (approximately below 0.15) and can in most cases be dropped altogether.

At each step of the algorithm, we choose random vector uniformly distributed in input area *i* and then find the closest node (in terms of Euclidean distance) between that point and the input stimuli. We then update the winner node and move their nearby nodes (those with conceptual distance within the radius *r*).

Each method generates a graph with minimum number of crossing, minimize the area of the graph and generate an internally convex planar graph. We have some examples as we can see in figures 5,6 .

We compare between three important isues: CPU time, drawing graph area in grid, and average length of edges using SOM and ISOM agorithms. In Table(1), The training time of the network effect directly on CPU time. So, we note that CPU time of SOM agorithm is less than ISOM agorithm. in compare with ISOM method. See the chart in figure 7.

TABLE I.    CPU TIME, AREA, AND AVERAGE LENGTH OF EDGES

| Example | Nodes of Graph | CPU time | | Area | | Average Length | |
|---|---|---|---|---|---|---|---|
| | | SOM | ISOM | SOM | ISOM | SOM | ISOM |
| 1 | 9 | 0.0842 | 0.0842 | 0.5072 | 0.3874 | 0.0752 | 0.0645 |
| 2 | 16 | 0.0936 | 0.0936 | 0.5964 | 0.5455 | 0.0397 | 0.0363 |
| 3 | 25 | 0.1310 | 0.1310 | 0.6102 | 0.5572 | 0.0212 | 0.0213 |
| 4 | 36 | 0.1498 | 0.1498 | 0.6438 | 0.6007 | 0.0142 | 0.0143 |
| 5 | 49 | 0.1872 | 0.1872 | 0.6479 | 0.6010 | 0.0103 | 0.0099 |
| 6 | 64 | 0.2278 | 0.2278 | 0.6800 | 0.6314 | 0.0077 | 0.0076 |
| 7 | 81 | 0.2465 | 0.2465 | 0.6816 | 0.6325 | 0.0060 | 0.0059 |
| 8 | 100 | 0.2870 | 0.2870 | 0.6677 | 0.6528 | 0.0049 | 0.0048 |
| 9 | 144 | 0.3962 | 0.3962 | 0.6983 | 0.6872 | 0.0034 | 0.0034 |
| 10 | 225 | 0.5710 | 0.5710 | 0.7152 | 0.6943 | 0.0021 | 0.0021 |

In VLSI applications, the small size of chip and the short length between the links are preferred. The main goals in our paper that minimize the area of output drawing graph on drawing grid, and minimize the average length of edges.

We note that ISOM method is better than SOM method to minimize the area and the average length of edges. In our experiments if the nodes greater than 400 nodes the SOM method generate graph with many crossing edges but ISOM generate graph no crossing edges in many times we train the program and ISOM is successes in minimize the graph area in compare with the SOM method .

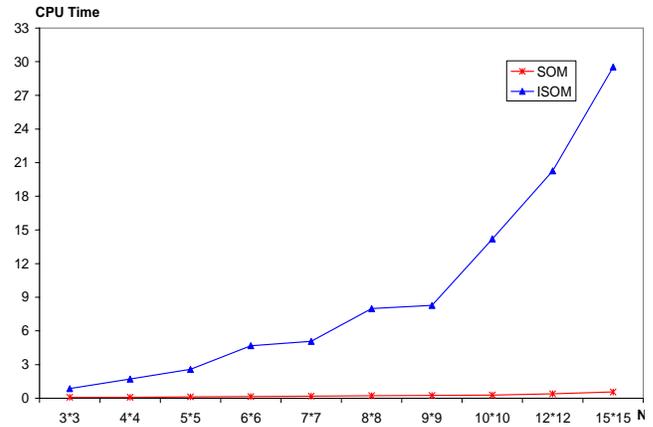

Figure 7.    Chart of CPU time using SOM and ISOM, respectively

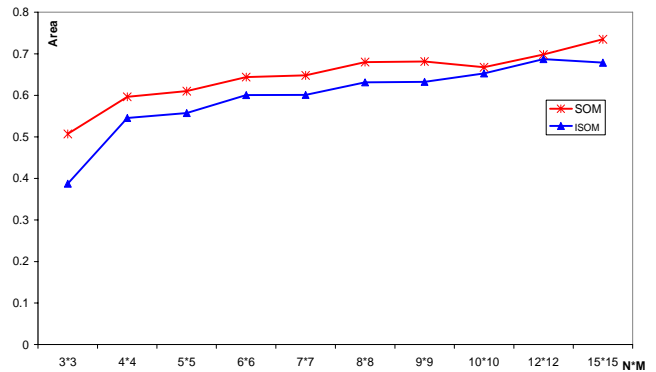

Figure 8.    Chart of graph area using SOM and ISOM, respectively

## VII.    CONCLUSIONS

In this paper, we have presented two neural network methods (SOM and ISOM) for draw an *internally convex of planar graph*. These techniques can easily be implemented for 2-dimensional map lattices that consumes only little computational resources and don't need any heavy duty preprocessing. The main goals in our paper that minimize the area of output drawing graph on drawing grid, and minimize the average length of edges which can be used in VLSI applications, the small size of chip and the short. We were compared between them in three important issues: CPU time, drawing graph area in grid, and average length of edges. We were concluded that ISOM method is better than SOM method to minimize the area and the average length of edges but SOM is better in minimize CPU time.

In future work we are planning to investigate three dimensional layout and more complex output spaces such as fisheye lenses and projections onto spherical surfaces like globes.